\documentclass[aps,showpacs,floatfix,nofootinbib]{revtex4-1}
\usepackage{comment}
\usepackage{xcolor}
\usepackage{graphics,graphicx,epsfig}
\usepackage[caption=false]{subfig}
\usepackage{amssymb,amsmath}
\usepackage{epstopdf}
\usepackage{amsmath}
\usepackage{mathtools}

\begin{document}

\title{Quantum synchronization and entanglement of indirectly coupled mechanical oscillators in cavity optomechanics: a numerical study}
\author{Devender Garg}
 \email{devender@iitrpr.ac.in}
 \author{Manju}
 \email{2018phz0009@iitrpr.ac.in}
\author{Shubhrangshu Dasgupta, and Asoka Biswas}%
\affiliation{%
Department of Physics, Indian Institute of Technology Ropar, Rupnagar, Punjab 140001, India
}%
\footnote{The first two authors contributed equally to this work.}
\date{\today}

\begin{abstract}
It is often conjectured that quantum synchronisation and entanglement are two independent properties which two coupled quantum systems may not exhibit at the same time. However, as both these properties can be understood in terms of the second order moments of a set of conjugate quadratures, there may exist specific conditions for simultaneous existence of entanglement and quantum synchronization. Here we present a theoretical scheme to achieve the same between two mechanical oscillators, which are indirectly coupled with each other via a coupling between two cavities. We show that in the presence of the cavity-oscillator coupling, quadratically varying with their displacements, these oscillators can be synchronized in the quantum sense and entangled as well, at times much longer than the decay time-scale of the cavity modes. Precisely speaking, we show that in the presence of quadratic coupling, entanglement criterion and quantum synchronization measure are simultaneously satisfied in steady state. This behaviour can be observed for a range of quadratic coupling, temperature, and frequency difference of the two oscillators. 
\end{abstract}

\maketitle

 \section{Introduction} \label{INTRO}
Two coupled harmonic oscillators, described by the displacements $q_j(t)$ from their respective equilibrium positions and the linear momenta $p_j(t)$ ($i\in 1,2$), are said to be synchronized, if they maintain $q_-(t)=\{q_1(t)-q_2(t)\}/\sqrt{2}\rightarrow 0$ and $p_-(t)=\{p_1(t)-p_2(t)\}/\sqrt{2}\rightarrow 0$, at long times. This corresponds to complete classical synchronization. The synchronization phenomenon was first observed by Huygens in the 17th century \cite{huygens}. A nice and detailed discussion of generic nature of synchronization can be found in \cite{pikovsky,balanov}.

In quantum systems, however, such synchronization is inherently not possible due to Heisenberg's uncertainty principle. 
As the position and momentum quadratures cannot be measured simultaneously at any time $t$, it is not possible to attain $q_{-}(t), p_{-}(t) \rightarrow 0$. In such case, the following quantification of synchronization has been proposed in terms of quantum fluctuations \cite{Mari}:
\begin{eqnarray}
S_{q}(t)&=&\frac{1}{\langle (\delta q_-(t))^2+(\delta p_-(t))^2\rangle}\; \label{sqm}\\
S_q(t) &\leq& \frac{1}{2 \sqrt{\left\langle \delta q_{-}(t)^{2}\right\rangle\left\langle \delta p_{-}(t)^{2}\right\rangle}} \leq 1,
\end{eqnarray}
where the angular bracket denotes the expectation value of the relevant operators and $\langle (\delta A)^2\rangle$ denotes the uncertainty of the operator $A$ ($A\in q_-, p_-$).
The uncertainty principle sets the value of $S_q$ between 0 and 1 and the complete quantum synchronization refers to the upper limit of the above inequality, i.e., $S_{q}=1$. Quantum-mechanical self-sustained oscillators have been shown to be synchronized to the external harmonic drive \cite{walter} or a different mechanical oscillator \cite{sadeghpour}. Marquardt and his coworkers have demonstrated that an array of optomechanical systems can exhibit synchronization \cite{Heinrich} that can be described by an effective Kuramoto-type model \cite{Kuramoto}. Quantum synchronization has also been studied in trapped ions, ensembles of atoms \cite{hush,xu1,xu2}, qubits, van der Pol oscillators \cite{zhirov,lee}, and Josephson junctions \cite{Xue,Nigg}.

One of the distinctive feature of quantum correlations is the nonlocality, that does not have any classical analogue. This refers to the possibility that measurement of one of the two coupled systems can affect the probability distributions of the other, which may be at a distance apart. There are various prescriptions of identifying nonlocality, including Bell's inequalities and EPR steering. To advocate for the local realism between two particles, Einstein, Podolsky, and Rosen (EPR) proposed that their position and momentum quadratures would maintain the relation $q_1=q_2$ and $\vec{p}_1=-\vec{p}_2$ at all times, as these paricles are coupled to each other in the stationary center-of-mass limit. In fact, in the classical sense, these particles may be considered synchronized. Moreover, the states of these two particles have been shown to exhibit nonlocality in terms of suitable Bell's inequality violation \cite{bellepr} and to correspond to an ideal quantum correlation - called EPR-correlation.

The local measurement of probability distributions is closely related to concept of separability. If a local operation on one of the subsystems changes the quantum probability distributions, then it clearly signals a quantum correlation. 
For a coupled bosonic system (like that of two harmonic oscillators), the joint variables $q_-$ and $p_-$ (which have been used to quantify synchronization in \cite{Mari}) satisfy the following uncertainty relation:
$ \langle(\delta q_-)^2\rangle+\langle (\delta p_-)^2\rangle \geq 1 $. If the state of this coupled system is separable, the following inequality is also satisfied, as shown by Duan and his coworkers (DC) \cite{Duan}:
\begin{equation}
\langle(\delta q_-)^2\rangle+\langle (\delta p_+)^2\rangle \geq 1\;, \label{dc}
\end{equation}
where the transformation $p_-\rightarrow p_+$ is made using partial transposition, using Peres's prescription \cite{Peres}.
This means that violation of this inequality (\ref{dc}) refers to inseparability. Such inseparable state is referred to being entangled. 

As clear from the above, both the quantum synchronization and the nonlocality are certain manifestation of quantum correlations and can be described or identified in terms of uncertainties of a set of joint EPR-like quadratures. The Heisenberg uncertainty relation has been used to suitably quantify the quantum synchronization as well as to establish conditions for inseparability (or, entanglement \cite{Simon}). Therefore, it is natural to expect certain interconnection between them. At least, there must exist certain regime of parameters, in which two oscillators can simultaneously exhibit quantum synchronization and entanglement. This could happen when $\langle(\delta q_-)^2\rangle \ne 0$ and  $\langle(\delta p_-)^2\rangle > \langle(\delta p_+)^2\rangle$. In this paper, we will explore a coupled-oscillator system in this regard.


We note that such interaction has been explored before, as well. For example, it was shown by Manzano {\it et al.\/} that two indirectly coupled oscillators, initially prepared in a separable state, can be both classically synchronized and entangled at long times \cite{Manzano}. This could be done by suitably tuning the coupling strength to the rest of the oscillators in the network. It was pointed out that synchronization helps in maintaining the entanglement even in presence of decoherence. How classical synchronization triggers the entanglement in many-body system has been studied in \cite{witt}. Lee and Cross have investigated both synchronization and quantum correlation between the two cavities with nonlinear crystals \cite{Cross}. They have shown in classical limit that two cavities can exhibit synchronization, while in the quantum limit, they get entangled. However, these works primarily focussed on classical synchronization.

To study quantum synchronization and entanglement of two harmonic oscillators indirectly coupled with each other, we can look forward to cavity optomechanical setup, in which a mechanical oscillator is coupled to a cavity mode. The mesoscopic oscillator can be quantized at low temperatures. So quantum-classical crossover can be studied in such a system, once the temperature is varied. In this paper, we will investigate if we can entangle such oscillators as well as quantum-synchronize them, at the same time.


In an optomechanical system, the leading order of coupling between the cavity mode and the mechanical mode is proportional to the displacement $q$ of the mechanical oscillator from its equilibrium position - a case of the so-called 'linear coupling'. That such a coupling can lead to cool the mechanical oscillator to its ground state using dynamical back action has been shown in \cite{Rae} and demonstrated in \cite{Arcizet,Chan}. This can also generate cavity-oscillator entanglement \cite{Vitali,Palomaki} and quadrature squeezing of mechanical mode \cite{Szorkovszky}. On the other hand, for specific configuration of the mechanical oscillators, when the quadratic dependence of the cavity-mechanical oscillator coupling on $q$ dominates, one can measure the energy eigenstate of the  mechanical mode \cite{Thompson}. Cooling and squeezing of the mechanical oscillator \cite{Nunnenkamp} and cavity-mechanical oscillator entanglement \cite{Liao2} have also been explored in such cases. It is further shown that quadratic coupling between a cavity mode and the motional degree of freedom of an atomic ensemble within the cavity can give rise to cavity nonlinearity at high probe laser power \cite{Purdy}.

To achieve only the quadrature coupling in optomechanical system, one requires to carefully place a membrane in the middle of the cavity, at one of the extrema of the cavity mode frequency. If such a constraint is relaxed, both the linear and the quadratic coupling can coexist in the same system. In such situations, one could also demonstrate squeezing of cavity quadratures \cite{Sainadh}, cooling of microspheres \cite{Xuereb} and of the mechanical oscillator to its ground state \cite{Rocheleau}, and self-sustained oscillations of mechanical oscillator \cite{Zhang}. In this paper, we show that to attain the quantum synchronization between two oscillators, along with entanglement, one requires to consider coexistence of both types of coupling, and that, in fact, the quadratic coupling enhances the synchronization.

The paper is organized as follows. In Sec. \ref{MODEL}, we present the model and derive the effective Hamiltonian. In Sec. \ref{Lang}, the relevant equations of motion are derived for the fluctuation terms. In Sec. \ref{Ent}, we present the relevant quantities and numerically investigate how both entanglement and synchronization can be simultaneously achieved between the oscillators. In Sec. \ref{con}, we conclude our paper.

\section{Model}   \label{MODEL}
We consider two optomechanical systems, in each of which a membrane is suspended inside a cavity. In addition, the two optical cavity (with resonance frequency $\omega_{cj}$, $j\in 1,2$) are directly coupled by an optical fiber between the inside mirrors, with a coupling constant $J$. We assume that the fundamental frequencies of the cavity modes are equal to $w_{cj,n} = n\pi c/L_j$, where $L_j$ is the length of the $j$th cavity $(j\in 1, 2)$, $n$ is a positive integer, and $c$ is speed of light in vacuum. The leading order of coupling between the cavity mode and the mechanical oscillator is proportional to the displacement of the oscillator from its equilibrium position.  We consider, in addition, a coupling quadratically varying with this displacement. One can achieve both orders of coupling in the membrane-in-the-middle setup, as we are considering, if the cavity frequency does not exhibit any extremum at the equilibrium position of the membrane \cite{bhatta}. That the suitable position and tilt of the membrane can generate both linear and the enhanced quadratic coupling has been demonstrated in experiments with optical cavity \cite{JHarris}.

The Hamiltonian of the system can be written as follows ($\hbar=1$):
\begin{eqnarray}\label{hamil0}
H_{ac} & = & H_0+H_g+H_i+H_p,\\
H_{0} & = &  \sum_{j=1,2} \omega_{cj} a_{j}^{\dagger} a_{j}+\frac{\omega_{m j}}{2}\left(q_{j}^{2}+p_{j}^{2}\right)\;,\nonumber\\
H_g & = &\sum_{j=1,2}\left[-g_{1}^{(j)} a_{j}^{\dagger} a_{j} q_{j}+ g_{2}^{(j)} a_{j}^{\dagger} a_{j} q_{j}^{2}\right]\;,\nonumber \\
H_i & = & -J(a_{1}^{\dagger}a_{2}+a_{2}^{\dagger}a_{1})\;,\nonumber \\
H_p & = & i E \left[ 1+\eta_{D} \cos\left(\Omega_{D}t\right)\right]\sum_{j=1,2} \left(a_{j}^{\dagger}e^{-i\omega_{lj}t }-a_{j}e^{i\omega_{lj}t }\right)\;,
\end{eqnarray}
where $H_{0}$ represents the unperturbed Hamiltonian, $a_j$ is the annihilation operator for the $j$th cavity mode, $q_j$ and $p_j$ refer to the dimensionless position and momentum operators for the $j$th mechanical oscillator, $H_{g}$ represents the optomechanical coupling between the mechanical oscillator and the cavity mode, $g_1^{(j)}$ and $g_2^{(j)}$ are the linear and the quadratic coupling  constant between cavity and mechanical resonator. They are defined as $g_1^{(j)}=\frac{\partial \omega_{cj}}{\partial q_j} x_{zpf}$ and $ g_2^{(j)}=\frac{\partial^{2} \omega_{ c j}}{\partial q_j^{2}} \frac{x_{zpf }^{2}}{2}$, where all the derivatives are calculated at the equilibrium position $q_j=0$. Here $x_{zpf}$ is the zero-point fluctuation of the mechanical oscillator's displacement given by $\sqrt{\frac{\hbar}{m_{j} \omega_{mj}}}$, where $m_j$ is the effective mass of the $j$th oscillator. The term $J$ is the coupling constant of cavity modes through an optical fiber. The driving of the cavity mode with external laser field of magnitude $E$ and frequency $\omega_{lj}$ is described via the Hamiltonian $H_p$. We have chosen that the driving fields are modulated with an amplitude $\eta_{D}$ at a frequency $\Omega_{D}$ .
\begin{figure}[h]
	\centering
	\includegraphics[width=.7\textwidth]{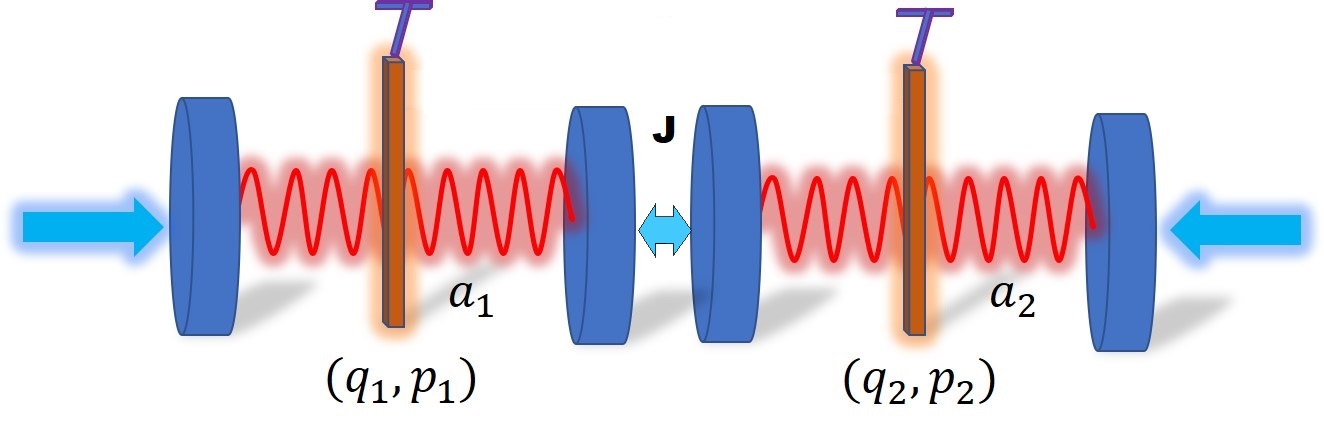}
	\caption{Schematic diagram of the two coupled cavity optomechanical systems. Each mechanical oscillator is suspended inside their respective cavity. The two cavity modes are externally coupled with each other via a coupling constant $J$.}
	\label{Model}
\end{figure}
In the rotating frame of laser frequencies, the total Hamiltonian $H_{ac}$ takes the following form:
\begin{eqnarray}\label{eq:H}
H & =& \sum_{j=1,2}\left[\Delta_{cj} a_{j}^{\dagger} a_{j}+\frac{\omega_{mj}}{2}\left(q_{j}^{2}+p_{j}^{2}\right)-g_{1}^{(j)} a_{j}^{\dagger} a_{j} q_{j}+g_{2}^{(j)} a_{j}^{\dagger} a_{j} q_{j}^{2}+i E \left(1+\eta_{D} \cos\left(\Omega_{D}t\right)\right)\left(a_{j}^{\dagger}-a_{j}\right)\right]-J(a_{1}^{\dagger}a_{2}+a_{2}^{\dagger}a_{1})\;,\nonumber\\
\end{eqnarray}
where $\Delta_{cj} =\omega_{cj}- \omega_{lj}$ is the detuning of the $j$th cavity mode from the respective driving field.

\section{Langevin equations}\label{Lang}
We will study the time evolution and the dissipation dynamics of the cavity and mechanical modes. From the above Hamiltonian the Langevin equations for the operators in Heisenberg picture can be obtained as follows:
\begin{eqnarray}\label{eq:p1}
\frac{d q_{j}}{d t} &=&\omega_{mj} p_{j}\;, \\
\frac{d p_{j}}{d t}&=&-\omega_{m j} q_{j}+g_{1}^{(j)} a_j^{\dagger} a_j-2 g_{2}^{(j)} a_{j}^{\dagger} a_{j} q_{j}-\gamma_{mj} p_{j}+\xi_j(t)\;, \\
\frac{d a_{j}}{d t}&=&-(\kappa_j+i \Delta_{cj}) a_{j}+i g_{1}^{(j)} a_{j} q_{j}-i g_{2}^{(j)} a_{j}q_{j}^{2}+iJa_{3-j}+E\left(1+\eta_{D} \cos\left(\Omega_{D}t\right)\right)+\sqrt{2 \kappa_j} a_{i n}^{(j)}\;,\nonumber\\
\label{eq:p2}
\end{eqnarray}
In the large mean field limit \cite{Aspelmeyer1}, we can expand the operators as a sum of their mean values and fluctuations as follows:
$a_j\rightarrow \alpha_j + \delta a_j$, $q_j\rightarrow {\bar q_j} + \delta q_j$, $p_j\rightarrow {\bar p_j} + \delta p_j$. From the above equations (\ref{eq:p1}-\ref{eq:p2}), the Langevin equations for the mean values can be obtained as follows:

\begin{eqnarray}
\frac{d \bar q_{j}}{d t} &=& \omega_{mj}\bar p_{j}\;, \\
\frac{d \bar p_{j}}{d t} &=& -\omega_{m j}\bar q_{j}+g_{1}^{(j)} |\alpha_j|^2-2 g_{2}^{(j)}  |\alpha_j|^2\bar q_{j}-\gamma_{mj} \bar p_{j}\;, \\
\frac{d \alpha_{j}}{d t} &=&-(\kappa_j+i \Delta_{cj}) \alpha_{j}+i g_{1}^{(j)} \alpha_{j} \bar q_{j}-i g_{2}^{(j)} \alpha_{j}\bar q_{j}^{2}+iJ\alpha_{3-j}+E\left(1+\eta_{D} \cos\left(\Omega_{D}t\right)\right)\;,
\end{eqnarray}
 The corresponding noise operators of above equations satisfy the following correlation relations for all $j$ \cite{Wall(2008)}:
 \begin{eqnarray}
 \left\langle a_{in}^{(j)}(t) a_{in}^{\dag (j)}(t')\right\rangle &= &  \delta(t-t')\;,\nonumber\\
 \left\langle\ a_{in}^{\dag (j)}(t) a_{in}^{(j)}(t')\right\rangle &= &   0 \;,\nonumber\\
 \left\langle\xi_j(t)\xi_j\left(t^{\prime}\right)\right\rangle&=&\frac{\gamma_{mj}}{2 \pi \omega_{mj}} \int \omega e^{-i \omega\left(t-t^{\prime}\right)}\left[1+\operatorname{coth}\left(\frac{\hbar \omega}{2 k_{B} T}\right)\right] d \omega\;,
 \end{eqnarray}
 where $k_{B}$ is the Boltzmann constant.
 The mechanical mode, coupled to the thermal bath, is affected by a Brownian stochastic force described by $\xi_j(t)$ with zero mean.
 The thermal bath is assumed to be a thermal equilibrium at a temperature $T$. For the case of large quality factor of the mechanical oscillator, the Brownian noise operator can be approximated in Markov approximation as
 $
 \left\langle\xi_{j}(t) \xi_{j}\left(t^{\prime}\right)\right\rangle=\gamma_{mj}\left( \bar{n}_{ m }+1\right) \delta\left(t-t^{\prime}\right)
 $,  where $\bar{n}_{m}=1 /\left[\exp \left(\hbar \omega_{mj} / k_{B} T\right)-1\right]$
 is the mean occupation number of the mechanical oscillators.

 Similarly, the Langevin equations for the fluctuations can be obtained as follows:
 \begin{eqnarray}
 \frac{d}{d t} \delta q_{j} &=& \omega_{m j} \delta p_{j}\;,\nonumber\\
 \frac{d}{d t} \delta p_{j} &=&-\omega_{m j} \delta q_{j}
 +g_{1}^{(j)}\left(\alpha_j \delta a_{j}^{\dagger}+\alpha_j^{*} \delta a_{j}\right)-2 g_{2}^{(j)} \bar q_j\left(\alpha_j^{*} \delta a_{j}+\alpha_j \delta a_{j}^{\dagger}\right)-2 g_{2}^{(j)}\left|\alpha_j\right|^{2} \delta q_{j} \nonumber \\
 & & -\gamma_{mj} \delta p_{j}+\xi_j(t)\;,\nonumber\\
 \frac{d}{d t} \delta a_{j} &= &-i[\Delta_{cj} \delta a_{j}-g_{1}^{(j)}\left(\bar q_j \delta a_{j}+\alpha_j \delta q_{j}\right)+g_{2}^{(j)} \bar q_j\left(2 \alpha_j \delta q_{j}+\bar q_j \delta a_{j}\right)]\nonumber \\
 & &-\kappa_j \delta a_{j}+iJ\delta a_{3-j}+\sqrt{2 \kappa_j} a_{i n}^{(j)}\;.\label{eq:f2}
 \end{eqnarray}

In the next section, we will show that the oscillators do not only get entangled, but also get quantum-synchronized. This result further strengthens the fact that quantum synchronization and entanglement are interlinked.

\section{Simultaneous Entanglement and Synchronization} \label{Ent}
It is known that the violation of the DC criterion is sufficient to detect entanglement for continuous-variable states.
Using standard algebraic identity, the following criterion for separability can be derived from the DC criterion:
\begin{equation}
E_D=\langle (\delta q_-)^2\rangle\langle (\delta p_+)^2\rangle\ge \frac{1}{4}\;,
\label{entcrit}
\end{equation}
violation of which indicates entanglement. This is the same criterion as derived by Mancini {\it et al.} for mesoscopic oscillators in optomechanical setup \cite{mancini}.
As was shown in \cite{mam}, the relation (\ref{entcrit}) is stronger than
the DC criterion. To investigate how the system get entangled with time, we will study the temporal dynamics of $E_D$ \cite{mancini}. 

Similarly, to investigate the quantum synchronization between two mechanical oscillators, we choose to investigate the behavior of the figure of merit $S_q$ given by Eq. (\ref{sqm}) \cite{Mari}.

To verify the above condition, we calculate the $S_q$ and $E_D$, by using the covariance matrix of fluctuations.
Let us start with the fluctuations equations (\ref{eq:f2}) in the following matrix form by defining the quadrature basis $\delta q_{aj}=\dfrac{(\delta a_j^\dag+\delta a_j)}{\sqrt{2}} $, $\delta p_{aj}=\dfrac{i(\delta a^\dag_j-\delta a_j)}{\sqrt{2}} $:
\begin{eqnarray}
\dot{ R(t)}=MR(t)+N(t)\;,
\label{eq:f3}
\end{eqnarray}
where $R(t)^T=\left( \delta q_{1},\delta p_{1},\delta q_{a1},\delta p_{a1},\delta q_{2},\delta p_{2},\delta q_{a2},\delta p_{a2}\right) $ and
\begin{eqnarray}
N(t)^T &=&\bigg(0,\xi_1,\sqrt{2\kappa_1}\delta q_{a1}^{in}, \sqrt{2\kappa_1}\delta p_{a1}^{in},0,\xi_2,
\sqrt{2\kappa_2}\delta q_{a2}^{in},\sqrt{2\kappa_2}\delta p_{a2}^{in}\bigg)
\end{eqnarray}
 is the noise vector. Here $\delta q_{aj}^{in}=(a_{in}^{(j)\dag} + a_{in}^{(j)})/\sqrt{2}$ and $\delta p_{aj}^{in}=i(a_{in}^{(j)\dag} - a_{in}^{(j)})/\sqrt{2}$, for $j = 1,2$.  The solution of this equation can be written as $R(t)=M(t)R(0)+\int^t_0 ds F(s)N(t-s)$, where $F(t)=\exp(Mt)$ and
\begin{equation}M=\left(\begin{array}{ll}
M_{1} & M_{0} \\
M_{0} & M_{2}
\end{array}\right)\;,\end{equation}
with
\begin{eqnarray}
M_j=
 \begin{pmatrix}
   0 & \omega_{mj} & 0 & 0 \\
-\omega'_{mj} & -\gamma_{mj} & \sqrt{2}G'_j  \operatorname{Re}\left(\alpha_j\right)& \sqrt{2}G'_j \operatorname{Im}\left(\alpha_j\right) \\
-\sqrt{2}G'_j  \operatorname{Im}\left(\alpha_j\right) & 0 & -\kappa_j & \Delta'_j\\
\sqrt{2}G'_j  \operatorname{Re}\left(\alpha_j\right)& 0 & -\Delta'_j& -\kappa_j
\end{pmatrix}\;,\nonumber\\
\end{eqnarray}
for $j = 1, 2$. The matrix $M_0$, as given below, describes the coupling between the oscillators:
\begin{equation}
M_{0}=\left(\begin{array}{cccc}
0 & 0 & 0 & 0 \\
0 & 0 & 0 & 0 \\
0 & 0 & 0 & -J \\
0 & 0 & J & 0
\end{array}\right)\;.\end{equation}
Here $G'_j=g_1^{(j)}-2g_2^{(j)}\bar q_j$, $\omega'_{mj}=\omega_{mj}+2g_2^{(j)}|\alpha_{j}|^2$, and  $\Delta'_j=\Delta_{cj}-g_1^{(j)}\bar q_j+g_2^{(j)}|\bar q_j|^2$\;.

To calculate the entanglement in any two subsystems, we can now calculate the covariance matrix $V(t)$, as a solution to the following linear differential equation:
\begin{eqnarray}
\dot{V}(t)=M(t) V(t)+M(t) V(t)^{ T }+D\;,
\label{eq:f4}
\end{eqnarray}
where the elements of $V$ can be identified as $V_{ij}= \left(\left\langle R_i(\infty)R_j(\infty)+R_j(\infty)R_i(\infty)\right\rangle \right)/2 $ and the diffusion matrix $D$ is given by
 \begin{eqnarray}
       D &=& {\rm diag}\big[
        0,(2n_m+1)\gamma_{m1} ,\kappa_1 , \kappa_1,
       0, (2n_m+1)\gamma_{m2}, \kappa_2 , \kappa_2 \big]\;,
        \label{dmat}
\end{eqnarray}

In the matrix $V$, every diagonal element represents the $2 \times 2$ matrix for the respective mode and every non-diagonal element $V_{ij}$ represents the $2 \times 2$ matrix of inter-mode covariance. 


The complete quantum synchronization $S_{q}(t)$ can be expressed in a concise form as
\begin{eqnarray}
S_{q}(t)  = \left\{\frac{1}{2}\left[V_{11}(t)+V_{55}(t)-V_{15}(t)-V_{51}(t)+V_{22}(t)+V_{66}(t)-V_{26}(t)-V_{62}(t)\right]\right\}^{-1}\;.
\end{eqnarray}
Similar expressions can be found for $E_D$ as well.

\begin{figure}
	\subfloat[\label{Q1Q21}]{%
		\includegraphics[height=5cm,width=.4\linewidth]{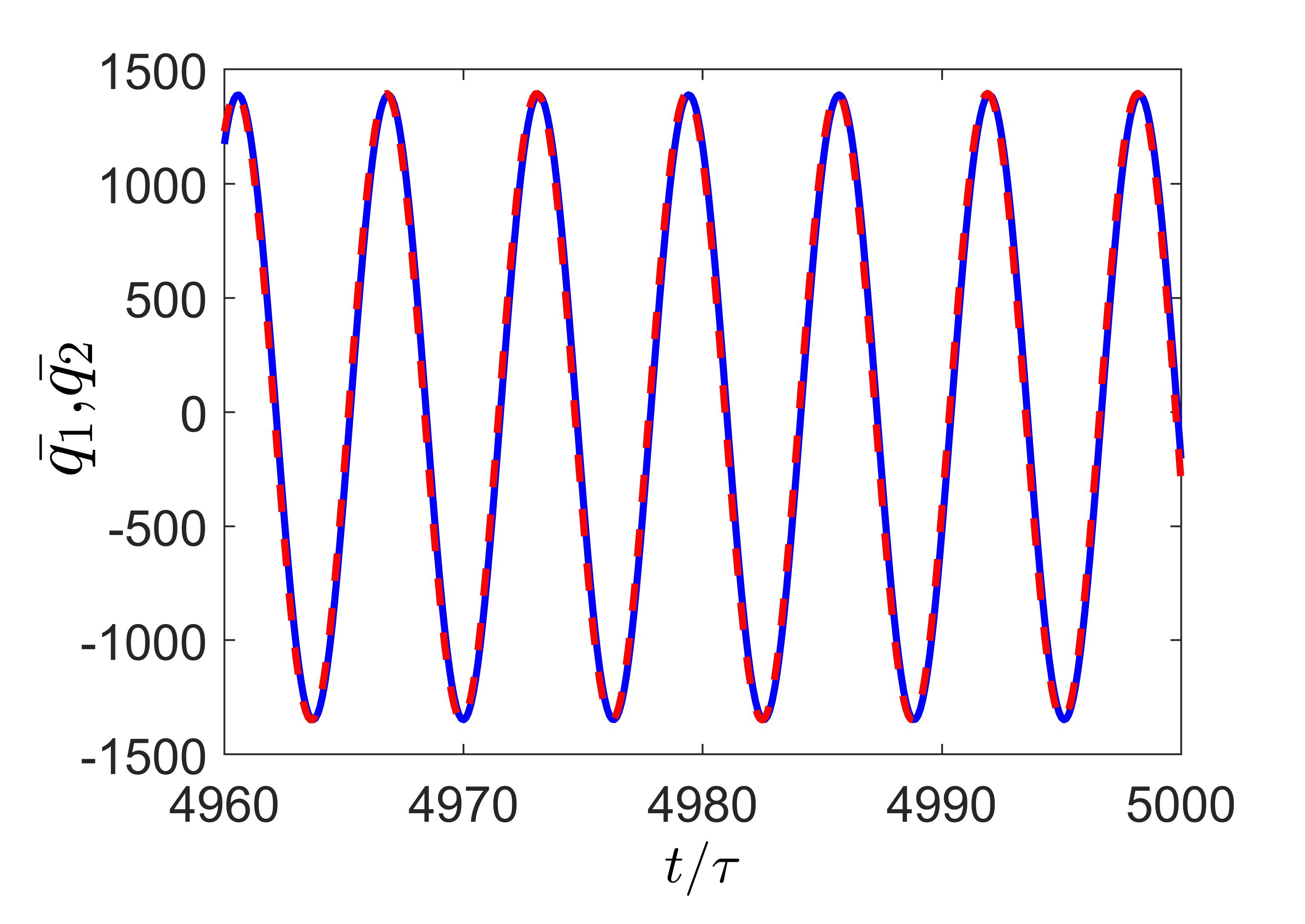}%
	}
	\subfloat[\label{P1P21}]{%
		\includegraphics[height=5cm,width=.4\linewidth]{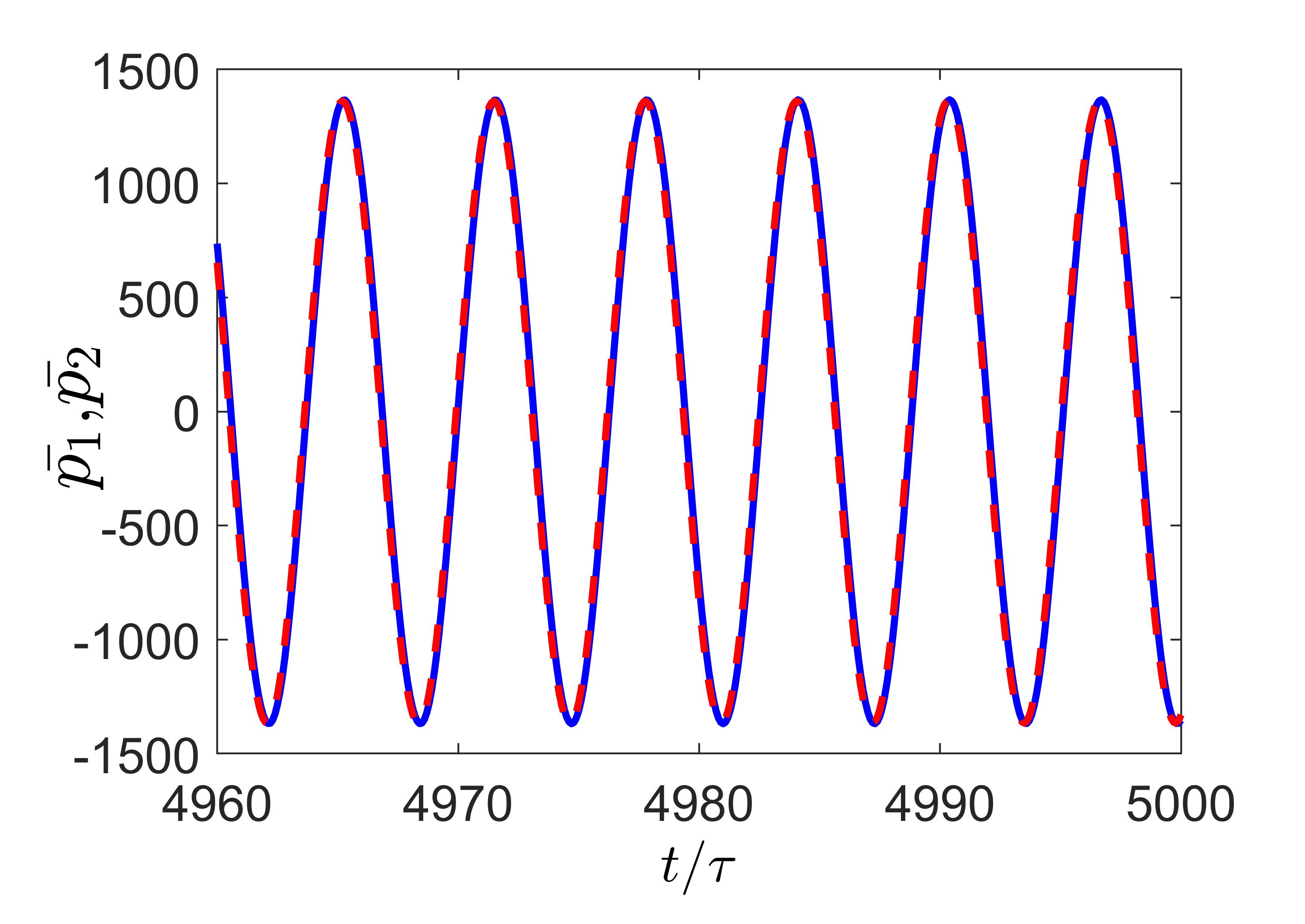}%
	}\\
	\subfloat[\label{SQ1}]{%
	    \includegraphics[height=5cm,width=.4\linewidth]{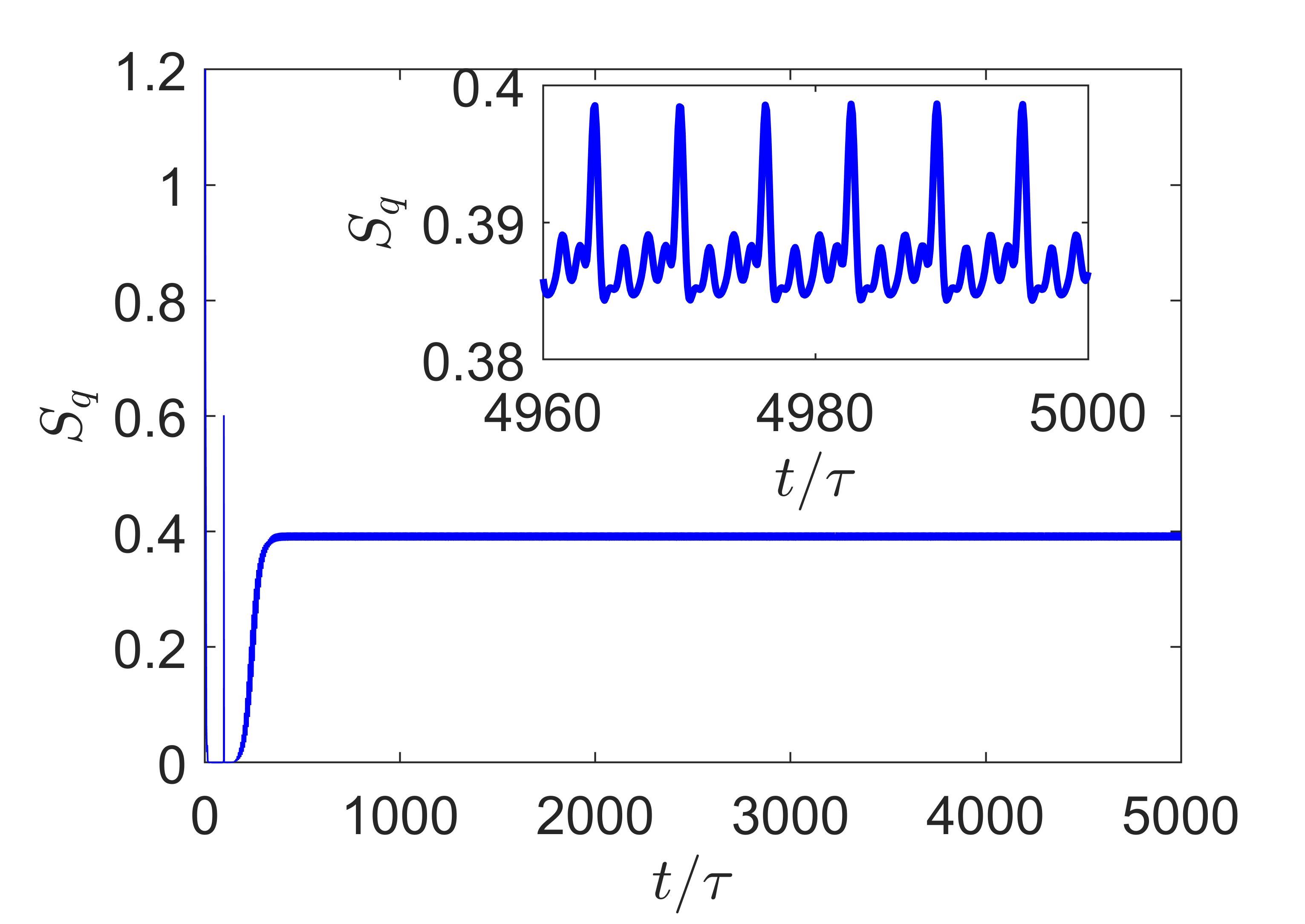}%
    }
    \subfloat[\label{ED1}]{%
        \includegraphics[height=5cm,width=.4\linewidth]{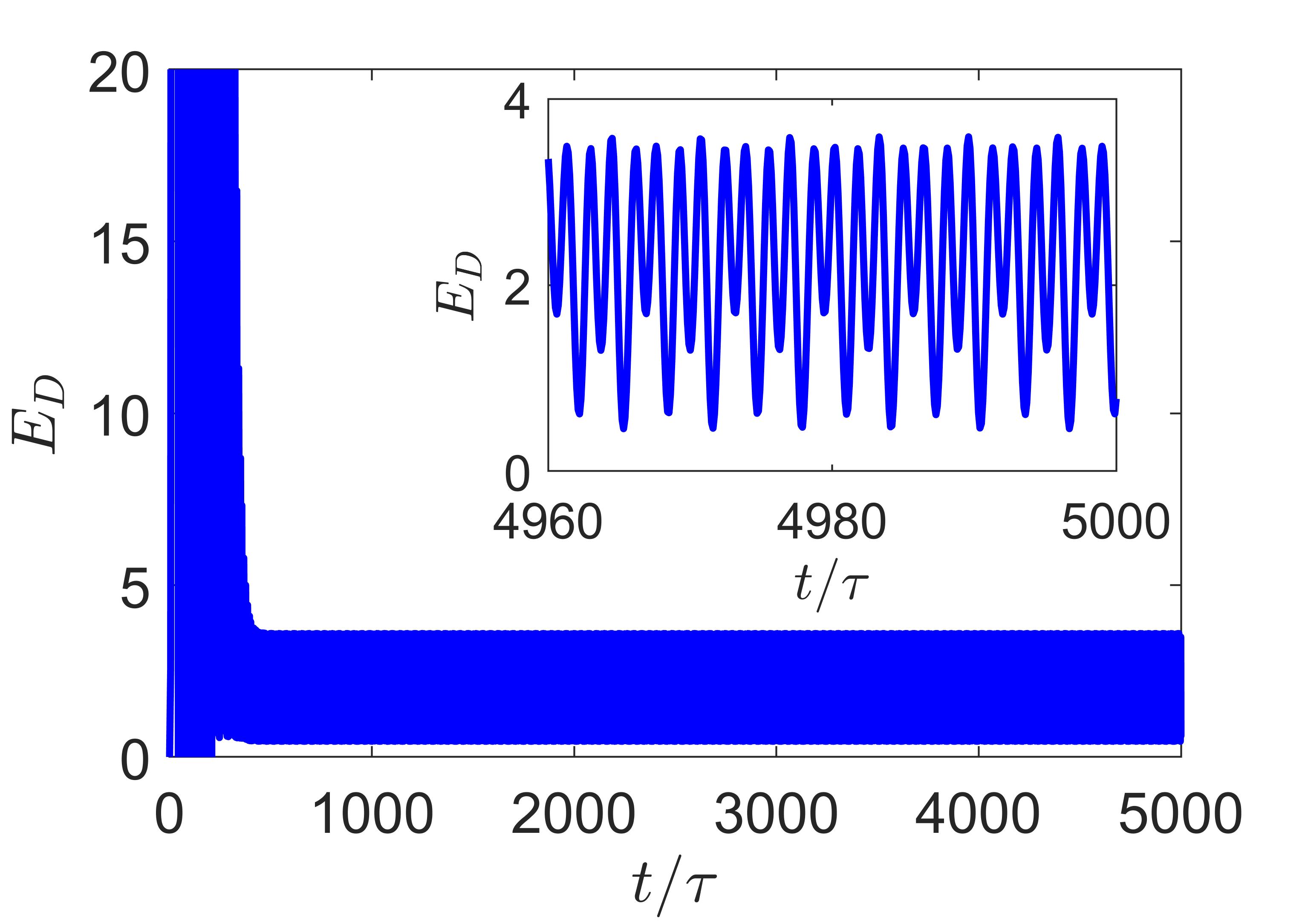}%
   }\\
 	\caption{Variation of (a) the mean values of $\bar{q_1}$ (red) and $\bar{q_2}$ (blue), (b) the mean values of $\bar{p_1}$ (red) and $\bar{p_2}$ (blue), (c) complete synchronization $S_{q}$, (d) entanglement $E_D$, with respect to time (in the units of $\tau=1/\omega_{m1}$).  Parameters chosen are $ \omega_{m1}=-\Delta_{c1}=1$, $\omega_{m2}=-\Delta_{c2}=1.005$, $T=0$ , $g_{1}=0.005$, $g_{2}=0$, $\gamma_{mj}=0.005$, $\kappa_j=0.15$, $E=100$, $\eta_{D}=1$, $\Omega_{D}=1$ and $J=0.04$. The inset shows the steady-state behaviour of the oscillation of $S_q$ and $E_D$.}
	\label{g20}
\end{figure}

\begin{figure}
	\subfloat[\label{Q1Q22}]{%
		\includegraphics[height=5cm,width=.4\linewidth]{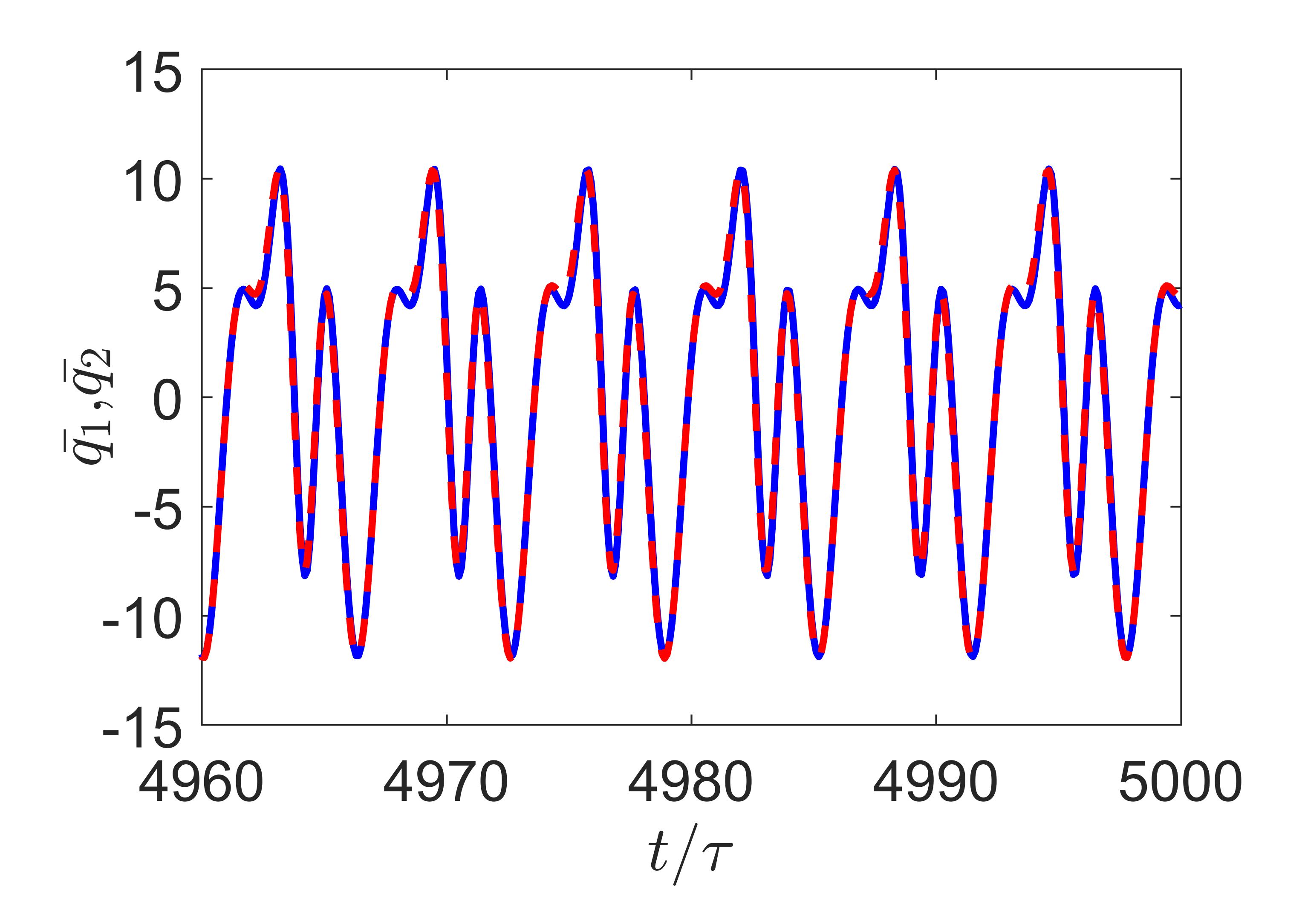}%
	}
	\subfloat[\label{P1P22}]{%
		\includegraphics[height=5cm,width=.4\linewidth]{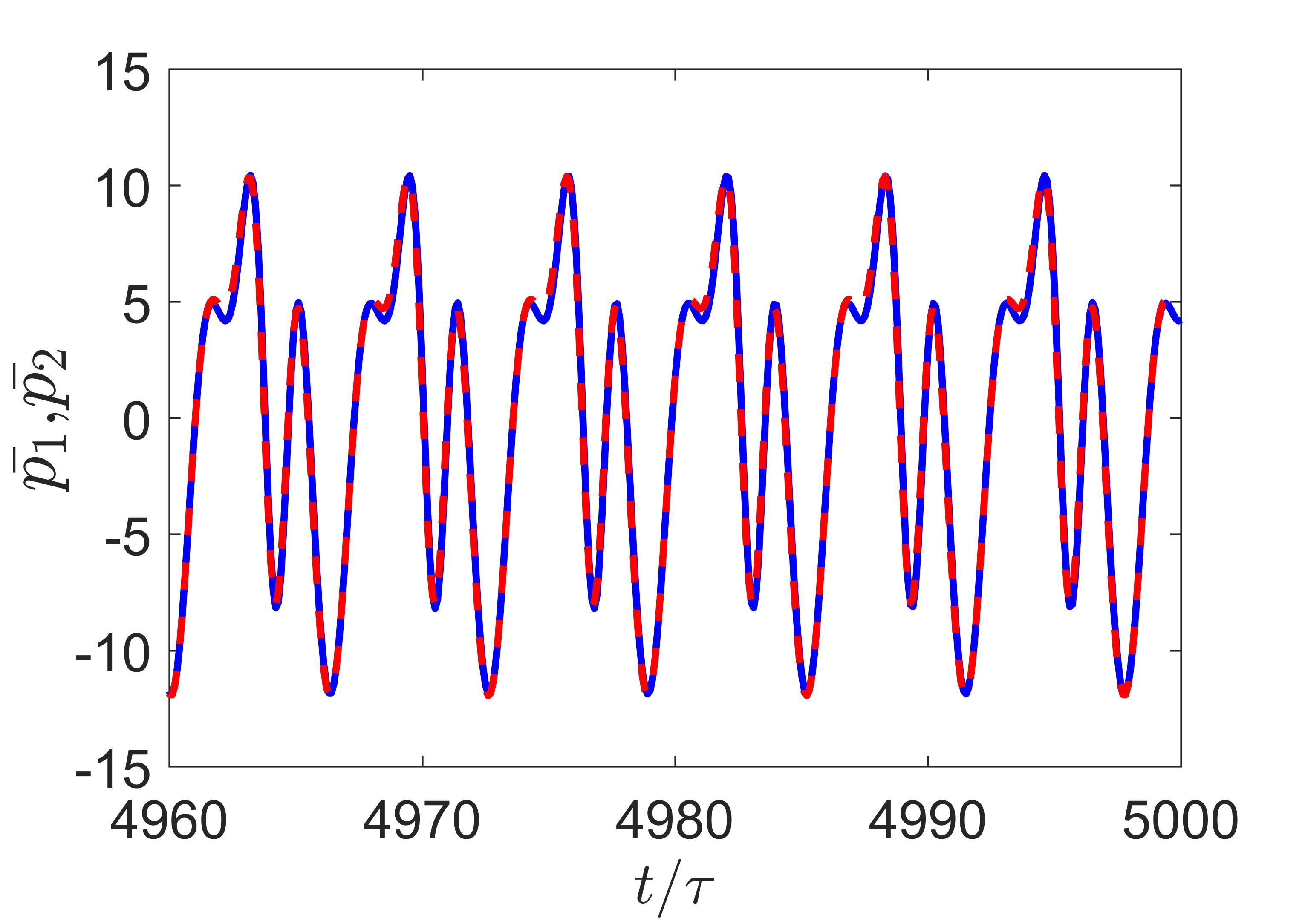}%
	}\\
	\subfloat[\label{SQ2}]{%
		\includegraphics[height=5cm,width=.4\linewidth]{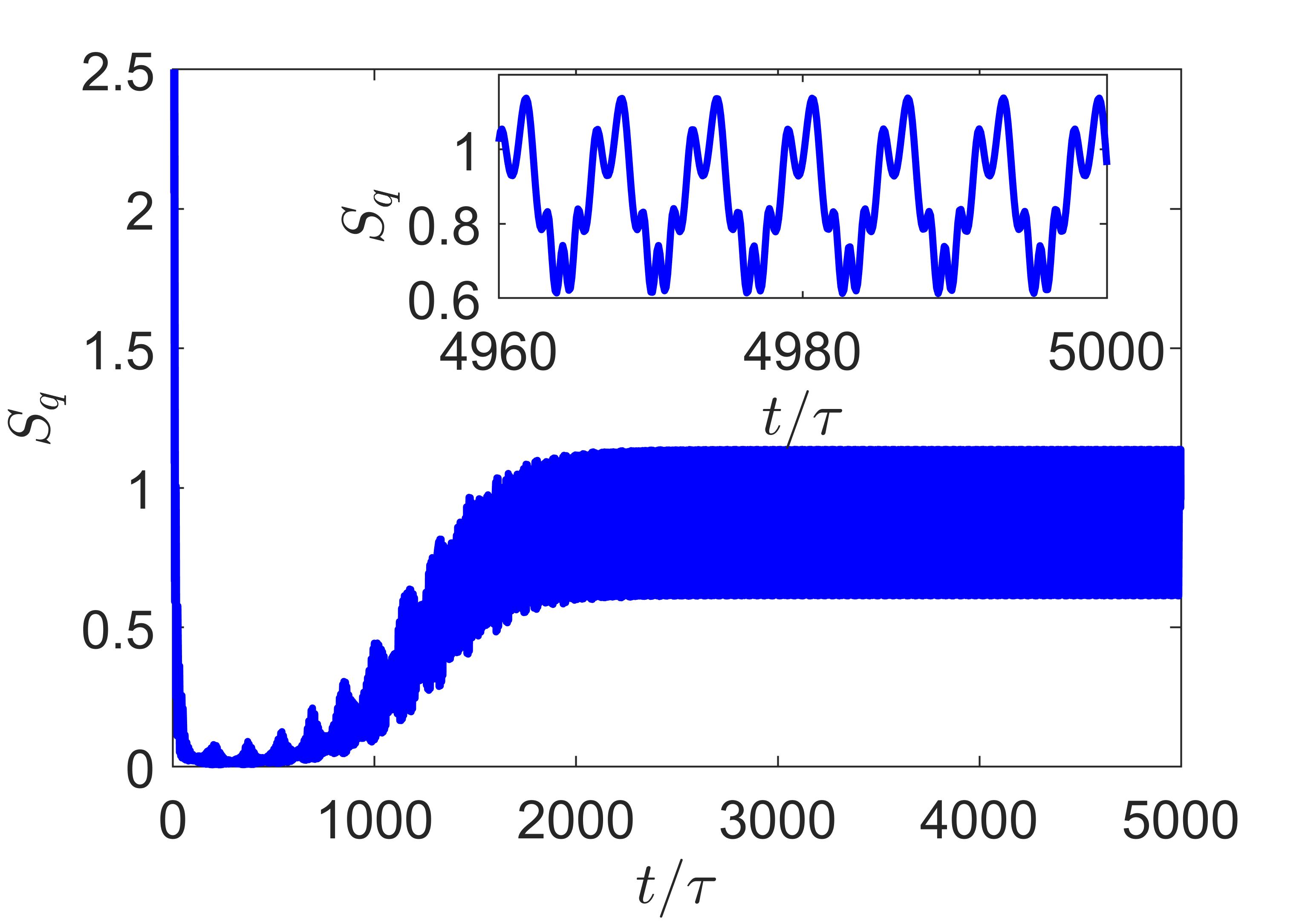}%
	}
    \subfloat[\label{ED2}]{%
    \includegraphics[height=5cm,width=.4\linewidth]{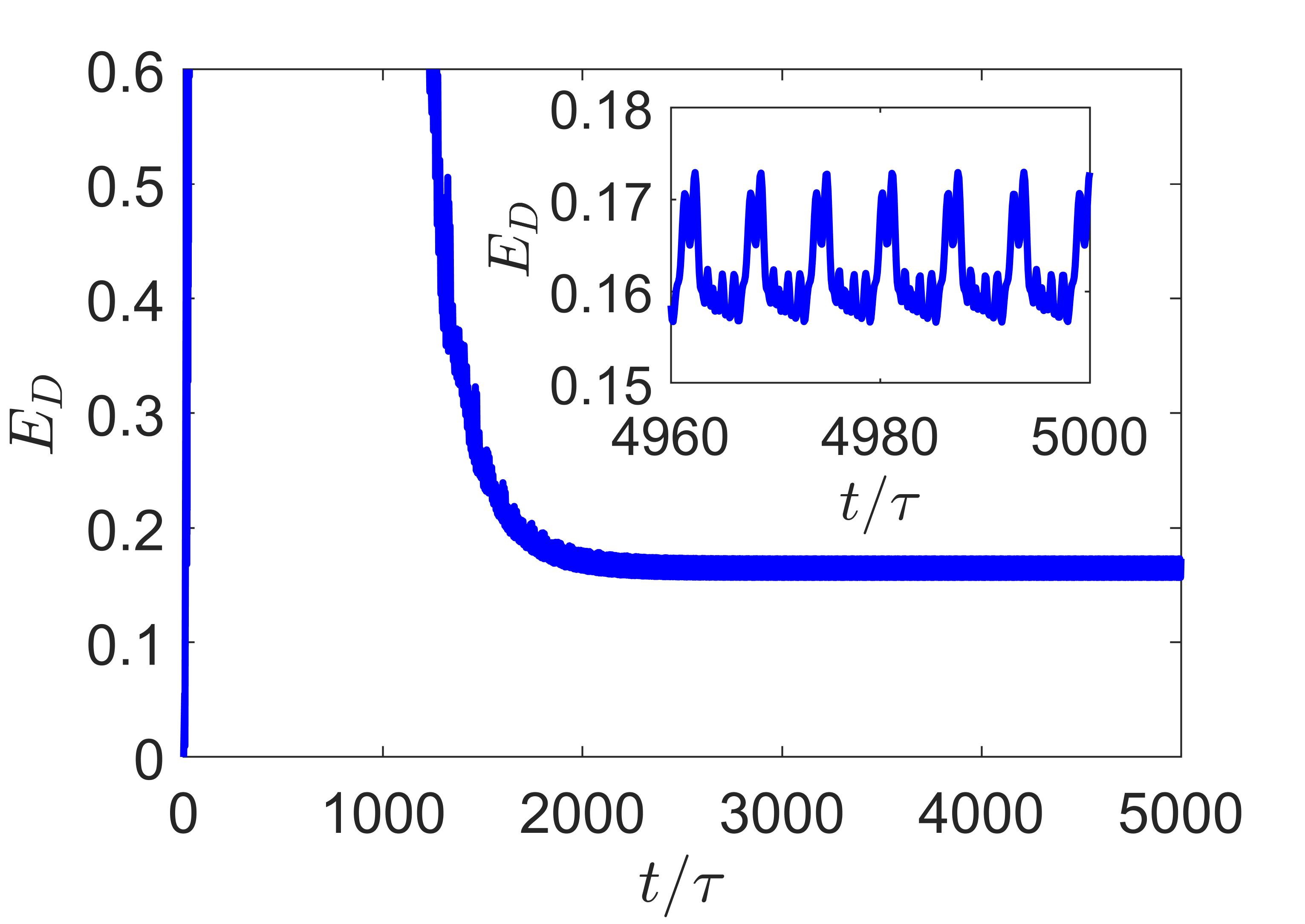}%
    }\\

	\caption{Variation of (a) the mean values of $\bar{q_1}$ (red) and $\bar{q_2}$ (blue), (b) the mean values of $\bar{p_1}$ (red) and $\bar{p_2}$ (blue), (c) synchronization $S_{q}$,  and (d) entanglement $E_D$, with respect to time (in the units of $\tau=1/\omega_{m1}$). Here we have chosen $g_{2}/g_{1}= 1\times10^{-2}$, while all the other parameters are the same as in Fig. \ref{g20}. The inset shows the steady-state behaviour of the oscillation of $S_q$ and $E_D$.}
	\label{Q1}
\end{figure}

\begin{figure}
	
	\subfloat[\label{nm}]{%
		\includegraphics[height=6cm,width=.45\linewidth]{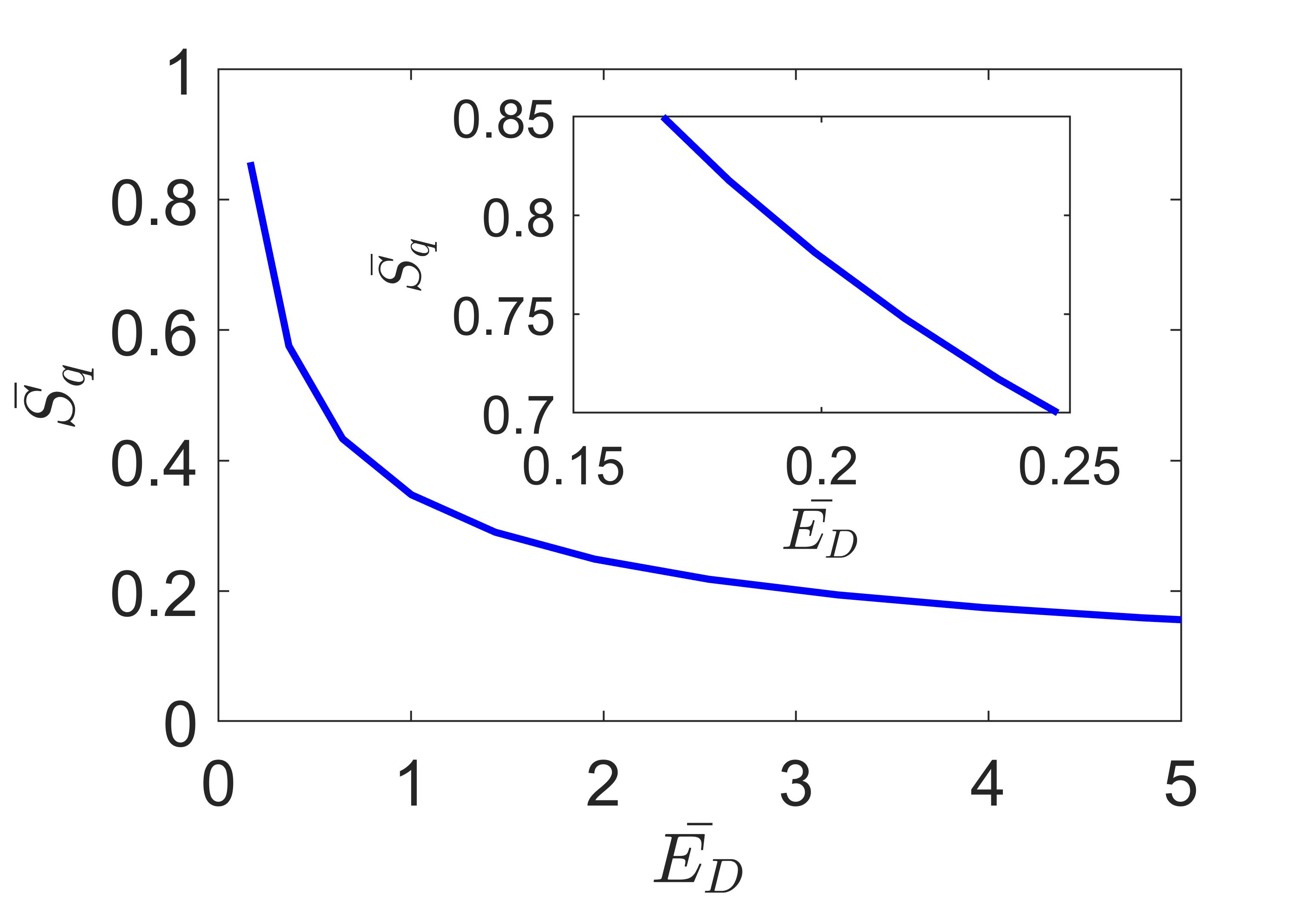}%
	}
	\hfill
	\subfloat[\label{Sqm}]{%
		\includegraphics[height=6cm,width=.5\linewidth]{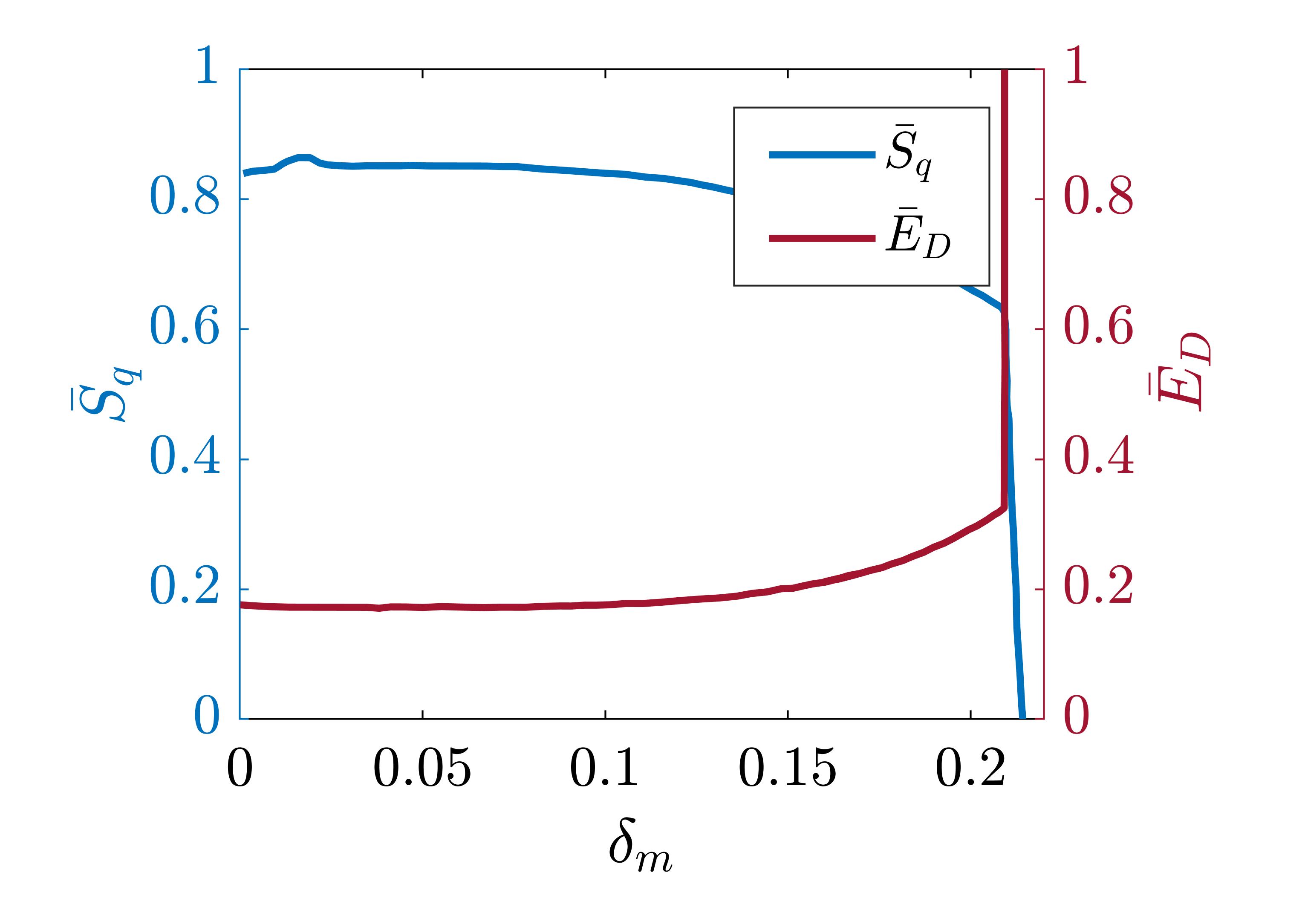}%
	}
	\hfill

	\caption{(a) A parametric plot of the variation of time-averaged values of synchronization $\bar{S}_q$ with the entanglement $\bar{E}_D$, when the number of thermal phonons $\bar{n}_m$ varies from 0 to 5. The inset shows the zoomed version of the same plot in the regions of parameters, when $\bar{E}_D\le 0.25$. (b) Variation of time-averaged values of synchronization $\bar{S}_q$ and entanglement $\bar{E}_D$ between the mechanical oscillators with respect to frequency difference $\delta_m = \omega_{m2} - \omega_{m1}$ of the mechanical oscillators. The other parameters are the same as in Fig. \ref{Q1}.}
	\label{wtemp}
\end{figure}



\subsection{Numerical Results and Discussion} \label{Dis}
In this section, we will show how this entanglement is built up with time and will investigate whether these oscillators get synchronized, as well. We choose here that the optomechanical couplings for both the mechanical oscillators are equal, i.e., $g_1^{(j)}=g_1$ and $g_2^{(j)} = g_2$ for all $j$.

We first consider the case when there is no quadratic coupling, i.e., $g_2=0$. As shown in the Figs.(\ref{Q1Q21}) and (\ref{P1P21}), the system will reach a  steady state at long times, when the mean values of the position and momentum vary periodically. This corresponds to onset of the limit cycles in phase space. In the Fig. (\ref{SQ1}) we show the time evolution of synchronization $S_{q}$ between the oscillators. The $S_{q} \sim 0.39$ at the steady state signifies existence on incomplete quantum synchronization. The small oscillations appearing in this figure are due to sinusoidal variation of the fluctuations in position and momentum at long times. It has been observed that there is no entanglement between the mechanical oscillators in this case of zero quadratic coupling. This is supported by the Fig. (\ref{ED1}), which shows that at long times, the separability criterion (\ref{entcrit}) is satisfied, as $E_D$ remains greater 0.25.

When we consider the quadratic coupling as well in our simulation, the synchronization increases beyond what was achieved without it. In the Figs. (\ref{Q1Q22}) and (\ref{P1P22}), we first show the time evolution of limit-cycle trajectories of the mean values $\bar{q_1}$, $\bar{q_2}$, $\bar{p_1}$, and  $\bar{p_2}$. This shows that the mechanical oscillators are classically completely synchronized. In the Fig. (\ref{SQ2}), we display the time evolution of quantum synchronization $S_{q}$. Clearly, even for a value of $g_2$ as small as $\sim 10^{-2}g_1$, the synchronization increases to a value close to unity and therefore the mechanical oscillators become nearly completely quantum synchronized. More importantly, the entanglement between the mechanical oscillators starts appearing with $g_2\ne 0$. The system also violates the standard inseparability criterion (\ref{entcrit}), as $E_D$ becomes less than 0.25 [see Fig. (\ref{ED2})]. A comparison between the Figs. (\ref{ED2}) and (\ref{SQ2}) clearly shows that the entanglement and synchronization are achieved, at a time scale of the similar order, and retain their values at long times, even in presence of decay [refer to Eqs. (\ref{eq:f2}) and (\ref{dmat})].

The numerical results for $g_2\ne 0$ suggest that one can get significant enhancement of quantum synchronisation and entanglement simultaneously between mechanical oscillators by choosing appropriate parameters. The direct coupling between the cavity (with a coupling constant $J$ ) and the linear coupling mediated by the cavity modes are proportional to $q_1$ and $q_2$, while the effective quadratic coupling between them varies as $q_1^2$ and $q_2^2$. Thus, the coupling proportional to $g_2$ imparts additional nonlinearity into the fluctuation dynamics, leading to the synchronization. That higher order terms are useful in obtaining synchronization has also been shown in \cite{witt}.

\subsection{Effect of indirect coupling and quadratic coupling}
From Eq. (\ref{eq:f2}), we can notice that the nonlinearity, arising due to coupling constant $g_2$, can suppress the oscillation of the two cavities as well as the photon exchange between them. Therefore, the oscillations of the positions and the momenta of the two mechanical oscillators are frozen, unlike the linear cases. This can be attributed to the opposite signs of the terms containing $g_1$ and $g_2$ in Eq. (\ref{eq:f2}). In this situation, the term containing $g_2$ tends to reduce the effect of that containing $g_1$ and the dynamics of the oscillators becomes more synchronized with each other (albeit with a $g_2$-dependent effective frequency). Therefore, the $S_q$ is enhanced and $E_D$ becomes less than 0.25, since the oscillation of the oscillators are suppressed, even for small values of $g_2$. Thus, it is the joint effect of the linear coupling and the quadratic coupling of mechanical oscillators with their respective cavity modes, that leads to simultaneous appearance of synchronization and entanglement between the oscillators. We emphasize that the quadratic coupling is essential to build up entanglement and to attain near-complete quantum synchronization for this particular optomechanical system.

Further note that for $g_2=0$, one can still have quantum synchronization, albeit not even near-complete and without entanglement.  It must be borne in mind that the classical synchronization (and the limit cycle) remains a necessary condition in any case. The above results are valid for a suitable direct coupling constant $J$. An interesting observation can be made when $J= 0$. In such a case, one cannot have any synchronization between the oscillators (and no limit cycle), as they are not coupled. Indeed, this has been previously noted by Mari {\it et al.\/} \cite{Mari}.

We show in the parametric plot Fig. \ref{nm}, how the synchronization and the entanglement change with the increase in thermal excitation. Clearly, the system remains maximally synchronized ($1-\bar{S}_q$ is minimum) when it maximally violates (\ref{entcrit}) [i.e., when $1/4 - \bar{E}_D$ is maximum, see Eq. (\ref{entcrit})]. This happens at absolute zero, i.e., for $\bar{n}_m = 0$. For larger excitation, the synchronization deteriorates and the entanglement decreases too, as $\bar{E}_D$ tends to $1/4$ (see  Figs. \ref{nm}). However, the synchronization remains more robust to temperature (or equivalently, the number of thermal phonons $\bar{n}_m$) as compared to the entanglement. Clearly, the entanglement vanishes (when $\bar{E}_D$ becomes larger than 1/4), for $\bar{n}_m\gtrsim 0.2$, while the system still maintains a residual quantum synchronization $\sim 0.15$ even for $\bar{n}_m$ as large as 5. We further show in the inset of the Fig. \ref{nm} the domain of simultaneous occurrence of synchronization and entanglement.   

We further show in Fig. \ref{Sqm} that with increase in frequency difference $\delta_m = \omega_{m2}-\omega_{m1}$, the synchronization and entanglement decreases in the similar way up to $\delta_m \sim 0.2$.  This means that both synchronization and entanglement are retained even when the frequency of the second oscillator, $\omega_{m2}$, is as large as 20\% more than that of the first oscillator, $\omega_{m1}$.


\section{Conclusion}\label{con}
In conclusion, we have presented a theoretical scheme to study the interplay between quantum synchronization and entanglement of two mechanical oscillators in a double-cavity optomechanical system. As both the criterion for entanglement \cite{mancini} and the measure of the quantum synchronization \cite{Mari} can be derived from the same uncertainty principle (in terms of their joint quadratures), we expected that there could be the possibilities of their simultaneous occurrence in the same system. In our model, each mechanical oscillator is suspended inside a cavity and is coupled with the cavity mode via a linear and a quadratic dependence on its displacement from the equilibrium position. Our numerical results shows that two coupled harmonic oscillators satisfy both the criteria at the same time and therefore, can be both quantum-synchronised and entangled simultaneously. To be more specific, appropriate choice of parameters in the presence of quadratic coupling can lead to greatly enhanced quantum synchronisation ($S_q > 0.85$) and entanglement between coupled oscillators, at a time much longer than the cavity decay time-scale. We have demonstrated classical synchronisation via limit cycle trajectories of the mean quadratures at long times, as a precondition to achieve quantum synchronization.We also investigated robustness of synchronization and the entanglement against thermal noise and frequency difference.

Though quantum synchronization can be considered arising out of certain quantum correlation, which could possibly be related to quantum discord \cite{Mari,sarma}, there can be certain parameter regime at which the quantum synchronization can also exist without entanglement (possibly, with nonzero discord). More importantly, the two phenomena (quantum synchronization and the entanglement) can be related to quantum fluctuations of a common set of quadrature variables (EPR variables). Precisely speaking, the entanglement criteria based on position and momentum quadratures can provide a suitable marker for quantum synchronization. As discussed in details in this paper, entanglement is associated with enhanced degree of quantum synchronization between two coupled oscillators. Note that entanglement refers to a nonclassical property of coupled bosonic systems, with a close equivalence to the nonlocality. Interestingly, any classical mixture with bosonic entangled state even exhibits quantum correlation, based on entanglement, as mentioned in the Introduction. Thus entanglement can stand, in its own merit, as closely related to quantum synchronization, as both arise out of quantum correlations in the quadratures. The existence of entanglement may be considered as a stronger criterion for near-complete quantum synchronization. We emphasize that we have put forward a common prescription to relate the two properties. The criteria that we have used can also be verified in experiments, based on quadrature measurements, contrary to quantum discord that cannot be directly verified in experiments.

We also emphasize that our results of quantum synchronization and entanglement are consistent with complete classical synchronization, as both $\bar{q_-}$ and $\bar{p_-}$ tend to zero, corresponding to a time-asymptotic limit cycle in classical phase space. Contrary to what Mari {\it et al.\/} conjectured, this is clearly not equivalent to mixed synchronization \cite{mixed} (that corresponds to $\bar{q_-}, \bar{p_+} \rightarrow 0$). Though the partial transpose (used to derive the Mancini criteria) refers to a local time-reversal (and hence counter-rotating trajectory in phase space), our results clearly do not involve any anti-synchronization (i.e., $\bar{p_+}\rightarrow 0$), as we have $\bar{p_-}\rightarrow 0$ instead.


\end{document}